# Wave propagation in filamental cellular automata

Alan Gibbons[1] and Martyn Amos[*,2]

[1] King's College London, Department of Computer Science,
London WC2R 2LS, United Kingdom
E-mail: amg@dcs.kcl.ac.uk

[2] Manchester Metropolitan University, Department of Computing and Mathematics,
Manchester M1 5GD, United Kingdom
E-mail: M.Amos@mmu.ac.uk

**Abstract**

Motivated by questions in biology and distributed computing, we investigate the behaviour of particular cellular automata, modelled as one-dimensional arrays of identical finite automata. We investigate what sort of self-stabilising cooperative behaviour these can induce in terms of waves of cellular state changes along a filament of cells. We discover what the minimum requirements are, in terms of numbers of states and the range of communication between automata, to observe this for individual filaments. We also discover that populations of growing filaments may have useful features that the individual filament does not have, and we give the results of numerical simulations.

## 1. Introduction

In both the realms of nanotechnology and natural biology there is the need to study how the behaviour of individual cellular components, acting through purely local stimuli, produce patterns of coordinated behaviour in extended systems made up of large numbers of these microscopic cells. *Cellular automata* have been successfully applied to the study of natural systems (Chopard & Droz, 1998; Deutsch & Dormann, 2005; Ermentrout & Edelstein-Keshet, 1993), and we are particularly interested in the emergence of oscillating *wave patterns*. Such patterns are central to the study of systems as diverse as cellular pattern formation (Alber, *et al.*, 2002), excitable media (Greenberg & Hastings, 1978) and physiological development (Koch & Meinhardt, 1994).

This paper reports on preliminary studies of particular cellular automata, namely one-dimensional strings (*filaments*) of identical finite automata (*cells*). A *filament state* is simply the string of states of the automata reading, say, from left to right along the filament. The automata take as input states of its neighbours and, depending on its current state, the input determines the next state of the automaton. Working in synchronised cycles this local behaviour determines successive states of the filament. We are then particularly interested in the behaviour, over time, of the filament state. Under what conditions may it exhibit coordination of the individual cellular components? For the domains of interest that we have in mind, we are interested in the simplest of automata, in terms of numbers of states, minimal input and design details that will induce coordination of cellular activity. In particular, we mainly concentrate on automata with no more than three states. Also, we only deal with the simplest of filaments, namely those consisting of *identical* finite automata.

## 1.1. Previous work

The classic synchronization problem in cellular automata is the so-called *firing squad problem* (introduced by Myhill in 1957, but not described in print until 1962 (Moore, 1962)). This concerns a line of identical finite automata ("soldiers"), each initialised to the same state (except for a single "captain" at far left). Soldiers take input only from their immediate neighbour(s), and the problem is to find a set of rules such that all soldiers enter a unique firing state at the same time. No 4-state solution to this problem exists, and the best-known solution has 6 states (Mazoyer, 1988). Although this is a classical problem, it is of tangential interest to us, because it seeks to home in on a one-off event, rather than looking for cyclical behaviour of the system. Others have been interested in similar issues. For example, Dijkstra (Dijkstra, 1974; Dijkstra, 1986) has discovered the existence of self-stabilising rings of automata. A self-stabilising system always returns to some "legitimate" configuration, no matter how it is perturbed (Burns & Pachl, 1989). Dijkstra described a solution to this problem for a ring of automata, where each machine may read the state of its two neighbours. However, Dijkstra's rings require more than one type of automaton for stable coordinated behaviour and, overall, his designs are more feasible for distributed digital computation, rather than the biological domain we have in mind. Others (Das *et al.*, 1995; Jiminez-Morales, *et al.*, 2002) have specifically concentrated on evolving finite automata for one-dimensional cellular automata arranged in a ring. These authors, in our opinion, entertain overly complicated designs of finite automata and their interest in rings imposes unnaturally on potential biological applicability. Also, it is not always clearly explained how an evolved automata induces the observed behaviour across the ring. Previous work in this area can all be classified in terms of the topology in the state space of the system, in which the (other) topology of the finite automata connections is *static*. We introduce a new dimension of *varying* this latter topology over time (by periodically extending filaments to simulate growth, at a pace slower than it takes for the system to stabilise) and this is entirely new with potential relevance to, for example, certain biological systems.

It is partly for these reasons that we have embarked on what we hope will be a systematic study and this paper reports some early findings. Not least amongst these is the discovery that very simple finite automata may induce stable coordinated behaviour in populations of filaments when the same automata are insufficiently powerful to be self-stabilising for individual filaments. Among other results, our studies also reveal a simple so-called *clock automaton* that, for any number of states, will increment the cellular states of the filament (*modulo* the number of states) in *unison*.

## 2. Some general points

Formally, we define a filament as a homogenous one-dimensional cellular automaton with specific non-periodic boundary conditions; namely, the state transitions of *end* cells in the lattice recognize an "empty" state representing missing cells in their neighbourhoods. That is, both the left-hand neighbour of the leftmost cell and the right-hand neighbour of the rightmost cell are assumed to be in this empty state. We take coordinated cellular behaviour to be signalled by sustained waves of cellular state changes along a filament. In a finite system, indefinite activity has eventually to be cyclic and this cyclic behaviour will be evident through wave patterns in the filament state. We study finite automata that induce these waves. There are essentially two kinds of waves that will be of interest:

- *Type A* waves in which a small number of cells change their state within each time step. These cells localise the wave front in space and at that particular time;
- *Type B* waves in which every cell changes its state within each time step. In general, Type A waves will be of greater interest. Cellular automata that sustain this type of wave are less susceptible to failure in the notion that there is synchrony in their action.

We need to indicate which finite automata we regard as *interesting*. These have state diagrams that are strongly connected and have minimum out-degree that is greater than one. In other words, there is a path (or sequence of inputs) that will take the automaton from any one state to any other, and in a given time step each state is not constrained (whatever the input) to be succeeded by one particular state. The former requirement is natural in terms of wishing each state to play a part. The latter requirement insists that each state change is *non-oblivious*; that is, that it depends on input and is not independent of it. Automata which have one or more states that can only be succeeded by a specific other state we call *oblivious automata*. Such automata can induce pathological cyclic behaviour. Consider the 2-state oblivious automaton that in each time step changes its state *whatever* this is and *whatever* the input.

For any initial filament state, the effect of this automaton is to immediately induce the cyclic behaviour whereby the state alternates between its initial description and that obtained by replacing each 0 by a 1 and *vice versa*. We would be right to regard such *Type B wave* behaviour as uninteresting. Not all the states have to be oblivious within an oblivious automaton to induce uninteresting cyclic behaviour in a filament. Consider the example of Table 1. Here the symbol * denotes any in the set {0, 1, φ}, where φ is the empty state which automata at the ends of the filament read from their absent neighbour. For this automaton, the 1 state is oblivious because it is *always* immediately succeeded by the state 0 whatever the input. For *any* initial filament, this automaton produces cyclic behaviour in which, at each step, the state of each individual cells alters. Three temporal traces of the automaton described in Table 1 are shown in Figure 1.

Within the filament state no string of 0's or 1's is longer than 2 and the appearance of strings of length 2 depends on the initial distribution of strings of 0's and 1's in the initial filament state. We do not take up space here justifying this description, but just state that the resulting wave patterns are too pathological to be of interest within our context (that is, the behaviour obtained is not "useful" or relevant in a biological context, since there is no observable sychrony).

| *Current* | 0 | 0 | 0 | 0 | 0 | 0 | 0 | 0 | 1 |
|---|---|---|---|---|---|---|---|---|---|
| *Input* | 0,0 | 0,1 | 0,φ | 1,0 | 1,1 | 1,φ | φ,0 | φ,1 | *,* |
| *Next* | 0 | 1 | 0 | 1 | 1 | 1 | 0 | 1 | 0 |

Table 1: Oblivious automaton (all possible transitions).

For the reasons described above, we disregard oblivious automata. It cannot be denied though that, in some cases and perhaps with a larger vocabulary of states and inputs, they do perform computations that, in a different context, may prove useful.

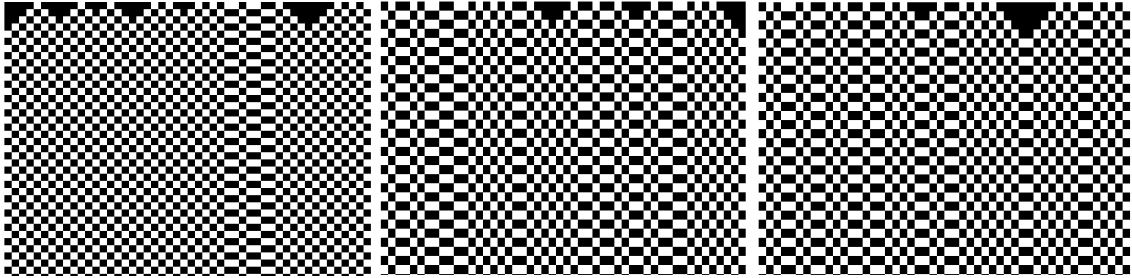

Figure 1: Temporal traces of oblivious automaton.

## 3. Type B waves and clock automata

Non-oblivious cellular automata, exhibiting Type B wave behaviour, all operate by individual cells copying (from left-to-right, say) the state of their neighbour, or some function of the state of their neighbour. We illustrate the principle in the Lemma below. The associated state of the cellular automaton exhibits a *standing wave* in which each cell, *in unison* with the others, counts the time *modulo* its number of states. Automata that induce this behaviour in filaments we call *clock automata*. Non-standing Type B waves are also clearly possible by the copying from left to right principle; their waveforms depend on both the permutation of states generated by the leftmost machine and the copying function employed by the others. By a *self-stabilising* automaton, we mean an automaton that will bring the filament to some prescribed cyclical behaviour *whatever* the initial filament state happens to be.

**Lemma 1.** *There exists a self-stabilising, clock automaton with any prescribed number of states.*

*Proof.* Table 2 describes the construction of the automata for 2 states. The table gives the next state of the automaton, for each possible input and its current state. Each input is in the form of a pair of states, the first is the state of automaton's left-hand neighbour and the second is the state of its right-hand neighbour. Here, φ denotes the *empty* state meaning that a left- or right-hand neighbour is absent because the automaton in question is at an end of the filament. The automaton works by making each internal and rightmost cell have, as its next state, ((*the state of its left-hand neighbour*) + 1) *modulo* 2. The leftmost cell simply increments its state by 1 *modulo* 2 at each time step. In this way, the state of the leftmost machine is essentially copied from left to right across the filament. It is trivial to prove, by induction on *i*, that after the *i*th time step ($i < n$) all cells within a distance *i* of the leftmost cell have the same state as the leftmost cell. Thus, self-stabilisation is achieved after *n* steps, where *n* is the length of the filament.

| Current | 0 | 0 | 0 | 0 | 0 | 0 | 0 | 0 | 1 | 1 | 1 | 1 | 1 | 1 | 1 | 1 |
|---|---|---|---|---|---|---|---|---|---|---|---|---|---|---|---|---|
| Input | 0,0 | 0,1 | 0,φ | 1,0 | 1,1 | 1,φ | φ,0 | φ,1 | 0,0 | 0,1 | 0,φ | 1,0 | 1,1 | 1,φ | φ,0 | φ,1 |
| Next | 1 | 1 | 1 | 0 | 0 | 0 | 1 | 1 | 1 | 1 | 1 | 0 | 0 | 0 | 0 | 0 |

Table 2: State transfer table for a 2-state clock automaton (all possible transitions).

Clearly such an automaton can be constructed for any number of states by replacing the next state, for each internal and rightmost automaton, by (*the state of its left-hand neighbour*) + 1) *modulo s*, where *s* is the number of states. In addition, of course, the leftmost cell simply increments its state by 1 *modulo s* at each time step. There is also the dual clock automaton which copies from right to left.

In Figure 2 we show several examples of the temporal evolution of the clock automaton.

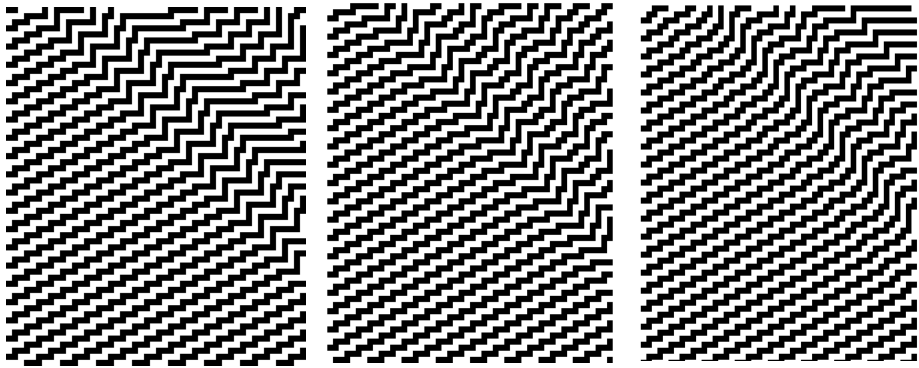

Figure 2: Three traces of the clock automaton.

In (Jiminez-Morales, *et al.*, 2002), the authors evolve a somewhat more complicated 2-state automaton to achieve the same outcome as our clock automaton but on a ring of cells. Our clock automaton works not only through the *copying from left to right principle* but also because the leftmost cell acts as a *source* of the wave, and the rightmost cell as a *sink*.

## 4. Type A waves with 2-state and 3-state automata

### 4.1. 2-state automata

Because of an exhaustive check of 2-state automata undertaken by us, we can *categorically* say that there exists *no* 2-state, non-oblivious finite automaton, *taking input only from its two immediate neighbours*, which generates a Type A wave. However, the following theorem shows that the situation changes if the range of inputs is extended.

**Theorem 1.** There exists a 2-state finite automaton that, taking input from its four nearest neighbours, induces a self-stabilising Type A wave.

*Proof (sketch).* Table 3 shows an important *subset* of the current states and inputs, ie. those which cause the finite automaton to *change* its state. Using this table it is easy to see that starting from the filament state in which the leftmost cellular state is 0 with all the others cells in state 1 ($[01^{n-1}]$), then the single 0 moves to the right (through a sea of 1's) until the rightmost end of the filament is reached. Here the single 0 is joined by a second and the pair, 00, then move to the left until the left hand end of the filament is reached. Here one of the 0's is lost and the initial filament state is again attained. The process is repeated indefinitely, and is depicted in Figure 3.

Concerning starting from a random initial filament state, additional rules ensure that strings of 0's of length greater than 1 move to the left while those of length 1 move to the right as above. Also, when string of 0's collide, they coalesce. At the left hand end of the filament all but one of 0's in a string are absorbed. Eventually a single 0 remains in the filament state and this places it within the normal cycle described above.

| Current | 0 | 1 | 1 | 1 | 0 | 0 | 0 | 1 | 1 | 0 | 0 | 1 | 0 | 1 |
|---|---|---|---|---|---|---|---|---|---|---|---|---|---|---|
| Input | 11,11 | 10,11 | 10,φφ | 11,00 | 10,φφ | 10,1φ | 10,11 | φ1,00 | φφ,00 | φφ,11 | φ0,11 | φ0,11 | φ1,11 | 10,1φ |
| Next | 1 | 0 | 0 | 0 | 1 | 1 | 1 | 0 | 0 | 1 | 1 | 0 | 1 | 0 |

Table 3: Subset of state changers for a self-stabilising 2-state automaton.

The cellular automaton defined by Table 3 generates perpetual cyclic behaviour using a different principle from clock automata. Instead of the end cells acting as source and sink, we see them here acting as reflectors of the wave. Taking input up to two cells away allows the process to occur.

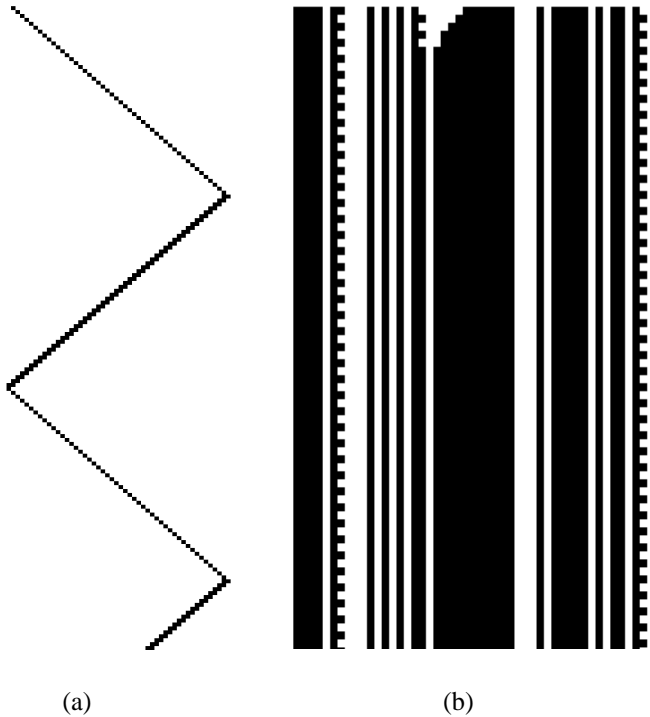

(a)                      (b)

Figure 3: Trace of self-stabilising 2-state automaton on (a) filament ($[01^{n-1}]$), (b) random filament.

## 4.2. 3-state automata

We know of *no* non-oblivious 3-state automaton, taking inputs only from its nearest two neighbours, that induces *self-stabilising Type A* waves in filament states. It is obvious though, from the previous section, that *self-stabilising Type A* wave generators must exist for automata which take their inputs from an extended range. Dijkstra (Dijkstra, 1986) described self-stabilising behaviour for a ring of 3-state finite automata, but his device required three different types of automata within the same ring. He believed that there is no such device which employs just one type of automaton.

We use this section, however, to introduce single 3-state automata that induce stable behaviour for *populations* of filaments undergoing individual growth (extension). This we do by first describing the behaviour of Automaton-I of Table 4, also shown in Figure 4.

| Current | 0 | 0 | 1 | 1 | 2 | 2 |
|---------|---|---|---|---|---|---|
| Input | {*,1} | {φ,1} | {*,2} | {φ,2} | {*,φ} | {φ,0} |
| Next | 1 | 2 | 2 | 0 | 0 | 1 |

Table 4: State transfer table for Automaton-I (only state-changing transitions).

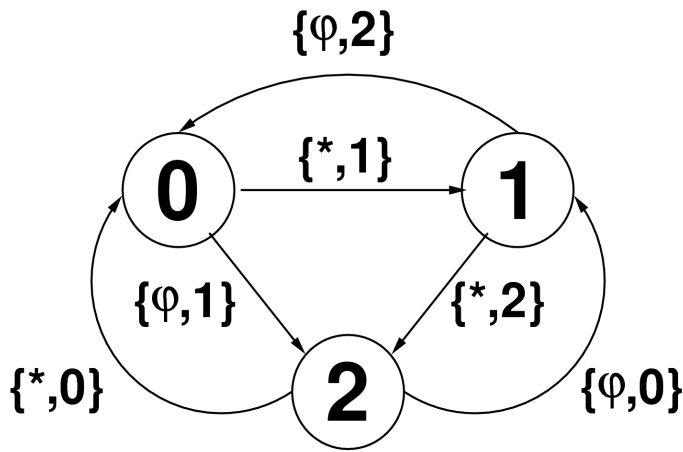

Figure 4: Graphical depiction of Automaton-I (only stage-changing transitions shown).

Table 4 records only those input pairs for a given state which lead to a state *change*, all other input pairs for that state do not cause a change of state. An input pair records the states of the neighbours of the automaton. The symbol * denotes any member of the set {0, 1, 2}. Again, the symbol φ denotes the *empty state* for cells at the end of the filament to denote the absence of one neighbour. This particular automaton is *symmetric*, in that the order of the states in the input pair is irrelevant because for any state change the same next state would be reached if the states of its neighbours were interchanged. This is why (in the interests of brevity) set brackets have been employed.

We now describe how Automaton-I can induce a Type A wave. Suppose that at one moment the system state is [0222 …. 2]. That is, the leftmost cell of the filament is in state 0 and all the rest are in state 2. We write this as $[02^{n-1}]$. It is easy to see that system states are then successively $[0^i 2^{n-i}]$ for $0 < i < n$. That is, a wave of 0's flows from left to right in the state description until the state $[0^{n-1}2]$ is reached. The next series of system states are successively $[0^{n-i}1^i]$, $0 < i < n$, as a wave of 1's flows from right to left. Then, in turn, a wave of 2's moves from left to right, a wave of 0's moves from right to left, a wave of 1's from left to right, and finally a wave of 2's moves from right to left when the state $[12^{n-1}]$ is reached. The next state is $[02^{n-1}]$, which is where we started, and the whole process repeats again and again indefinitely (Figure 5(a)). This pattern is characterised by there only ever being one cellular state change in each system state change and therefore, as long as the system remains within this cycle of behaviour, we do not necessarily require the cells to operate in simultaneity.

Automaton-I is not *self-stabilising*. There are initial filament states that do not take us into cyclic behaviour (see, for example. Figure 5(b)). Inspired by our belief that no 3-state self-stabilising automata exits that, taking input only from its two nearest neighbours, induces Type A wave behaviour, we are motivated to study further machines like Automaton-I which clearly induce the same cyclic behaviour for many different, but not all, initial filament states. This is the subject of the following section.

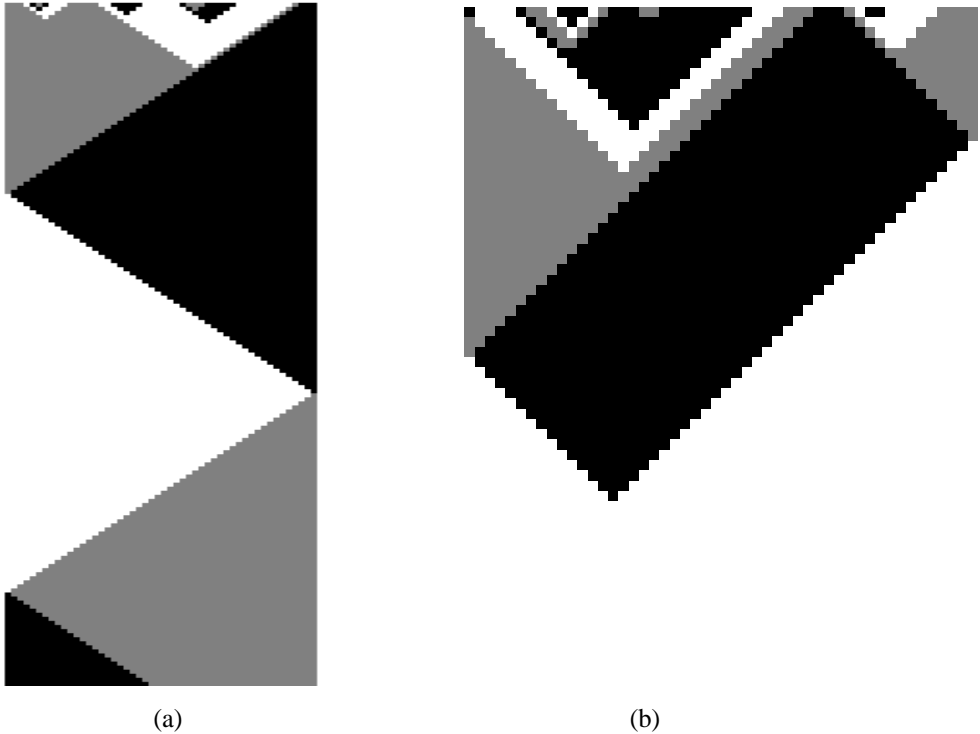

(a)            (b)

Figure 5: (a) Cyclic behaviour. (b) Non-cyclic behaviour.

## 5. Viable populations of 3-state Symmetric Automata

Here we discover a novel phenomenon. It is that populations of growing filaments of certain finite automata may induce persistent behaviour which the same automaton cannot induce in all the individuals starting from random filament states.

In order to introduce this, we return to Automaton-I of Table 4. Given a filament state, we define a *step* to be a change in states as we pass between adjacent cells. We further say that the wave-like behaviour, described in the previous section, is *normally cyclic*. The only filament states that induce no cellular state changes are $[0^n]$, $[1^n]$ or $[2^n]$ and these states we call *quiescent*. We then have the following Lemma.

**Lemma 2.** *For Automata-I type filaments, any initial filament state containing an odd number of steps*, in $O(n)$-*time, becomes normally cyclic and all other initial filament states become quiescent.*

*Proof (sketch).* The state changes of the automaton are so designed that, wherever a step occurs in the system state, then the step moves by adding 1 (modulo 3) to that cellular state which will bring the two cellular states into equality. This is the mechanism that causes the *wave* that characterise the earlier description. The internal reflection of waves at the filament ends are caused by cellular state changes involving the φ symbol; these reverse the sense of a step. In that description, there was just one wave (or *step*) active any moment. As long as that is the case, the system will behave as described.

There are two other cases to consider: the existence of *no* steps in the system state and the existence of *two or more* steps. If there are no steps in the system state, then its description will be one of: $[0^n]$, $[1^n]$ or $[2^n]$ and the system state will remain unchanged indefinitely.

Now suppose that there are just *two* steps in the system state of our fixed length filament. Without loss of generality, we may assume that these two steps are moving towards each other (if they are not, then a later reflection of a wave at a filament end will ensure that this is so). The system state will then be one of the forms: $[0^i 2^j 0^k]$, $[1^i 0^j 1^k]$ or $[2^i 1^j 2^k]$ for some $i + j + k = n$ and $i, j, k > 0$. If it is not already, the value of $j$ will become 1 or 2. For both cases, the next system state will be one of $[0^n]$, $[1^n]$ or $[2^n]$. The two waves can be interpreted as having cancelled each other out.

This cancelling out of two waves is a principle that can now be used to explain the asymptotic outcome for any filament state which initially has more than two steps. It is not difficult to see that, over time, waves cancel themselves out in pairs until just one or no waves exist. If one exists, then the asymptotic behaviour will be *normally cyclic*. If no waves exist, then the filament will be *quiescent*.

Of the $3^n$ possible initial filament states of length $n$, how many have an odd number of steps (#odd($n$)) and how many have an even number (#even($n$))? An exact combinatorial analysis shows that:

for $n$ even, #odd($n$) = $(3^n + 3)/2$ and #even($n$) = $(3^n - 3)/2$
for $n$ odd, #odd($n$) = $(3^n - 3)/2$ and #even($n$) = $(3^n + 3)/2$.

For simplicity and brevity, we assume here that $n$ is very large and that all initial filament states are equally likely, so that the probability of an arbitrary initial filament becoming *normally cyclic* is then 50%. Given a *normally cyclic* filament, we note that an increase in its length by the accretion of a single cell (whose state is with equal probability 0, 1, or 2) will, with probability 2/3, introduce a step in the filament state which will then become *quiescent*. With probability 1/3 it will remain *normally cyclic*. Similarly, such an extension to a *quiescent* filament will, with probability 2/3 make it *normally cyclic* and, with probability 1/3, it will remain *quiescent*. These observations beg the following picture.

Imagine a population of filaments of the same size ($n$) which all grow, in a soup of free cells, by the accretion of single cells at the same regular interval. By the arguments above, we see that initially half the filaments are *live* (normally cyclic) and half are *dead* (quiescent). After an update of the population by cell accretion as described, we ask what proportion of the filaments (all now of size $n+1$) is now live and what proportion is dead? The proportion of live cells of length ($n+1$) is 1/3 of 1/2 (from the live cells of length $n$) plus 2/3 of 1/2 (from the dead cells of length $n$). That is, again, 50% of the cells are live and so 50% are dead. It is clearly a constant of the dynamics that, at any one time, half the population is live although the individual filaments making up this proportion constantly change. A little thought also shows that the arguments employed are robust in terms of dropping some of the assumptions (that all the filaments are the same length, that there is simultaneity in cell accretion and so on).

We coin the term *viable population* to describe a population of filaments which, in the process of growth, experiences live behaviour in a fixed proportion of its number of filaments but for which there is a continuous turnover of the individuals that make up this proportion. So the whole population contributes to the property that a fixed proportion is live and this contrasts with an individual filament that is either live or dead.

It seems to us likely that *viable populations* might have a place both in devices of nanotechnology and in biology. Such populations exhibit behaviour induced by automata which, we believe, are too simple to guarantee useful behaviour in individual filaments. It seems there is strength in numbers!

We have described one viable population and it is natural to ask if there are others. Table 5 depicts an alternative automaton, similar to Automaton-I (also represented graphically in Figure 6).

| Current | 0 | 1 | 1 | 2 | 2 |
|---|---|---|---|---|---|
| Input | *,1 | *,0 | φ,1 | *,0 | φ,0 |
| Next | 1 | 2 | 2 | 0 | 1 |

Table 5: State transfer table for Automaton-II (only state-changing transitions).

Here, the conventions adopted for Table 4 apply; that is, the Table only records those input pairs for a given state which lead to a state *change*. We say that filaments constructed from this automaton are of Automaton-II type. Such filaments, as is easily seen, have two types of *quiescent* state: (1) strings consisting only of 1's and 2'; (2) strings consisting of 0's only.

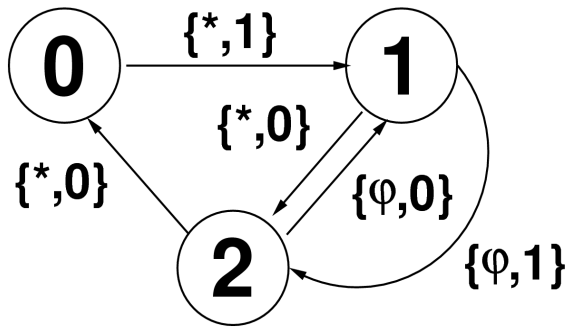

Figure 6: Graphical depiction of Automaton-II (only state-changing transitions shown).

We can also identify *normally cyclic* behaviour starting from the filament state $[0^{n-1}1]$. The next two states are $[0^{n-2}11]$ and $[0^{n-3}122]$, then there are successive states $[0^{n-i}12^{i-1}]$ for $3 < i < (n-1)$. The states then following $[012^{n-2}]$ are $[02^{n-1}]$ and $[0^j 2^{n-j}]$ for $1 > j > (n-1)$, the last of these being $[0^{n-1}2]$. The next state is $[0^{n-1}1]$ and so the cycle repeats. In brief, it consists essentially of a flow of 2's (headed by a single 1) from right to left followed by a flow of 0's from left to right. The cycle time is $2n$ compared with the $6n$ of type Filament-I. This is illustrated in Figure 7.

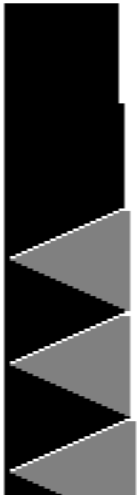

Figure 7: Trace of Automaton II run on filament $[0^{n-1}1]$ (black pixels represent state 0, white state 1, and grey state 2).

**Lemma 3.** *For Automaton-II type filaments, any initial filament state containing a 0 at one end and not containing 0 at the other, in O(n)-time, becomes normally cyclic and all other initial filament states become quiescent.*

*Proof (sketch).* If there is not a 0 at either end of the filament state, there is no way in which a 0 can get into such a position. This is because there are no state transitions of Automaton-II from states 1 or 2 which involve φ in the input and cause the next state to be 0. For such filament states, 2's adjacent to 0's become 0's and 0's adjacent to 1's become 1's, repeatedly, until the entire filament state consists of 1's and 2's only.

If there is a 0 at *one* end of the filament state but not at the other, then this remains the case. This is partly because there is no state change from state 0 involving φ. Without loss of generality assume that the 0 is at the left-hand end of the filament state. The leftmost part of the filament state is then a string of 0's (of length at least one). If the next symbol to the right is a 2, then the string of 0's expands until a 1 is encountered (it might merge with other 0's on the way). When a 1 is encountered then it is "dragged" to the left by a contraction of the string of 0's until it is removed by action near the filament end when the string of 0's again expands to the right (this is similar to that observed in *normally cyclic* behaviour). This activity repeats until the filament state is either $[0^i 2^{n-i}]$ for some $i < n$ or is $[0^{n-1}1]$. These states signify that the filament has become *normally cyclic*.

If there is a 0 at *both* ends of the initial filament state, then the strings of 0's at each end behave in just the same way as the string of 0's in the previous case except that here they finally collide somewhere in the interior of the filament the wave motion is cancelled out and the filament state becomes $[0^n]$.

The worst case, for any initial filament state to become *quiescent* or *normally cyclic*, is provided by the state $[01^{n-1}]$. The rightmost 0 in the string of 0's is easily seen to "travel" a distance of O($n$) in contractions and expansions of the string before the filament becomes *normally cyclic*. The complexity claim follows.

Of the possible initial filament states, it is easy to see that those without a zero at either end, those with a 0 at one end, those with a 0 at both ends occur in proportions respectively of 4/9, 4/9, 1/9. That is, a proportion of 4/9 of a uniform population of filaments will be *live* and a proportion of 5/9 will be *dead*. Now, if we allow the filaments to grow by the accretion of single cells as before, we see that: (1) of the filament states initially having a 0 at both ends, 1/3 retain that property and 2/3 have a 0 at one end only; (2) of those filament states initially having a 0 at one end and not at the other, 1/2 retain that property, 1/6 have a 0 at both ends and 1/3 have no 0 at either end; (3) of those filament states initially having no 0 at either end, 2/3 retain that description and 1/3 have a 0 at one end and not at the other.

It then follows that after such growth of the filaments, the proportions of the three types of filament state are retained and we see that uniform populations of Automaton-II type filaments, like those of Automaton-I, form *viable populations*. The following theorem holds.

**Theorem 2.** *Populations of Automaton-I and Automaton-II type filaments form viable populations. In the former case, the proportion of live filaments is (asymptotically) 1/2 and in the latter case 4/9. The normal cycle length for Automaton-I is 6n and an arbitrary initial filament state becomes quiescent or normally cyclic in at most O(n)-time. The normal cycle length for Automaton-II is 2n and an arbitrary initial filament state becomes quiescent or normally cyclic in at most O(n)-time.*

We believe that symmetric 3-state automata inducing *viable populations* are not rare, but more work is required to fully understand this.

## 5.1. Numerical simulations

We now briefly describe the results of numerical simulations carried out to investigate the population-level behaviour of filaments of both Automaton-I and Automaton-II. For each set of experiments, a population of $m$ filaments of length $n$ was created. The state of each filament was initially randomised, and then each was simulated for 5000 iterations. Crucially, filaments were allowed to periodically *grow*; after a set interval, the right end of each filament was extended by the addition of a randomly-initialised "cell".

We first describe the results for Automaton-I. We varied *m* between 5 and 200, and chose a growth interval of 6*n* (the normal cycle length). At each time step, we measured the proportion of "live" (i.e., active) filaments in the population.

In Figure 8, we notice spikes in the level of activity whenever filaments grow, as the addition of a cell can often "revive" previously dead filaments. For small population sizes, the proportion of live filaments varies considerably, but, as expected, stabilises at roughly 50% for larger (100-200 filaments) populations. Inspection of the simulation confirms the important observation that there is constant *turnover* in terms of which filaments are active at any one time.

We performed a similar set of experiments for Automaton-II, using different population sizes. The reason for this was the shorter normal cycle time (2*n* as opposed to 6*n*), which gave rise to "chaotic" population-level behaviour for small values of *m*. Again, we observed (Figure 9) that the proportion of live filaments roughly stabilised at the expected value of about $4/9 = 44\%$.

## 6. Summary and open problems

In term of the number of states and range from which they take input, we have described the simplest finite automata that, for filamental cellular automata, induce regular cyclic behaviour. In modes of behaviour where every cell changes its state in each time unit, the most interesting of these was the *clock automaton*. For, perhaps the more interesting mode of behaviour where a small number of cells are active at each time step, we described a self-stabilising 2-state automata. For cellular automata, consisting entirely of one type of finite automaton only, we know of no other machine with the same characteristics that is as simple. Finally, we introduced the novel notion of *viable populations* of filaments which exhibit stable characteristics under growth induced by automata that are not powerful enough to induce stable behaviour in individual filaments.

In terms of future work, we need to know more about *viable populations* in particular. Many automata will produce wave-like behaviour along filaments for a subset of possible initial filament states. But do all of these induce *viable populations* or only a subset? Also, we have dismissed oblivious *behaviour* as inducing uninteresting cyclic behaviour for our purposes. This may not be so easy to do for filaments with a richer vocabulary of states and range of inputs. The questions we pose for filaments here may, naturally be extended to cellular automata with more dimensions.

## Acknowledgement

The authors thank two anonymous reviewers for useful comments.

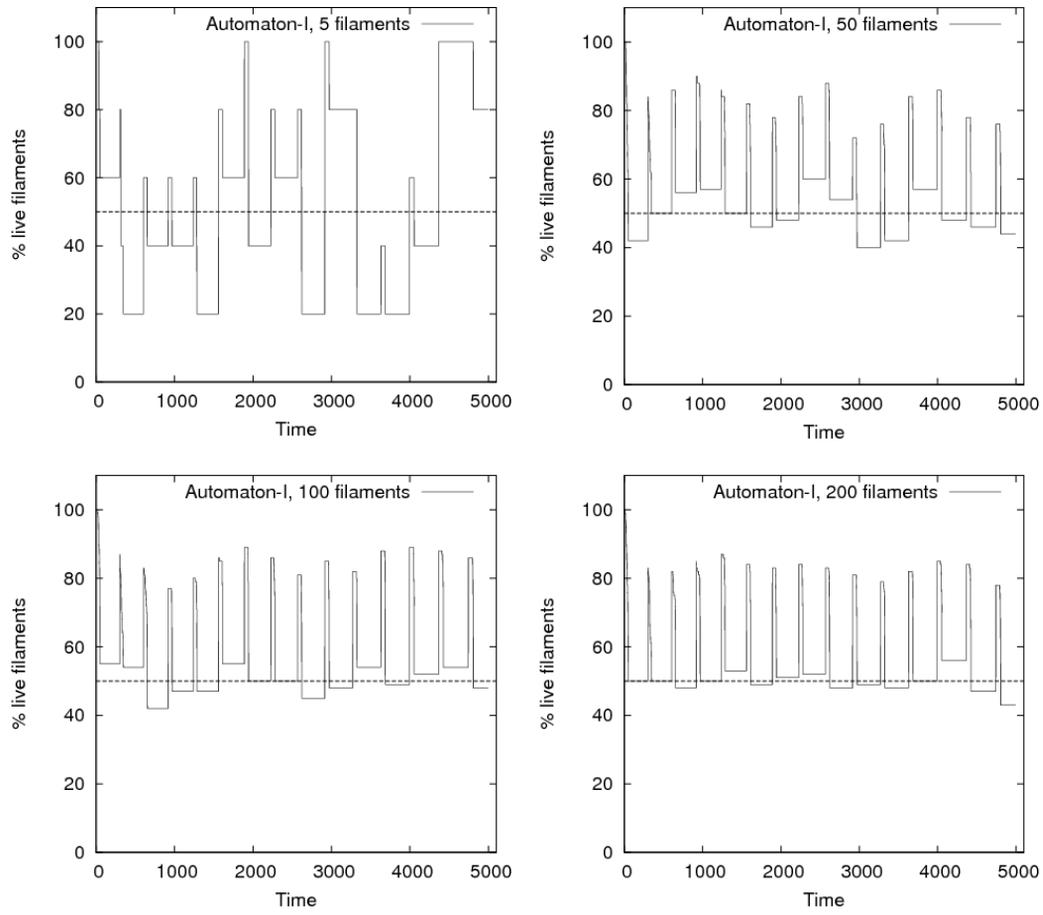

Figure 8: Results of numerical simulations of Automaton-I for 5, 50,100 and 200 filaments.

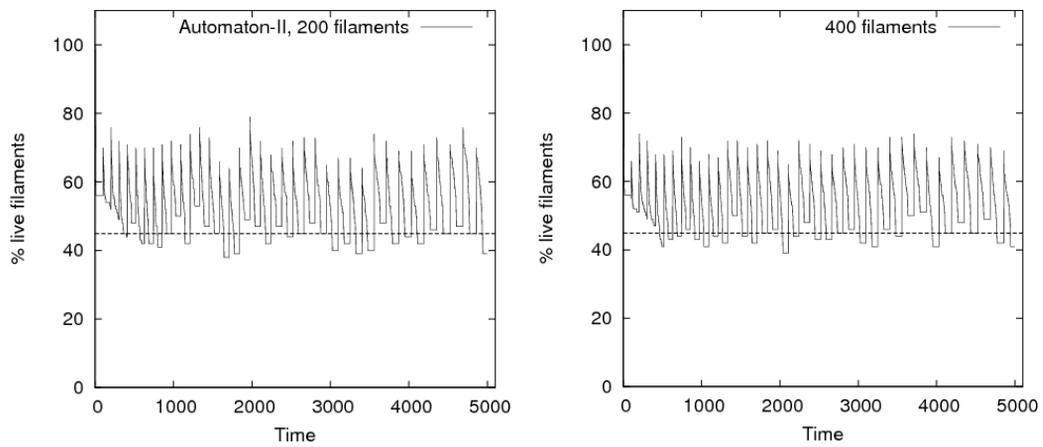

Figure 9: Results of numerical simulations of Automaton-II for 200 and 400 filaments.